\documentclass{elsart}

 \usepackage[dvips]{graphicx}
\usepackage{subfigure}

\usepackage{amssymb}
\usepackage{wasysym}

\begin{document}

\begin{frontmatter}

\title{HoloTrap: Interactive hologram design for multiple dynamic optical trapping}

\author{E. Pleguezuelos,}
\ead{encarni.pleguezuelos@ub.edu}
\author{A. Carnicer,}
\author{J. Andilla,}
\author{E. Mart\'in-Badosa,}
\author{M. Montes-Usategui}
\address{Universitat de Barcelona \\Departament de F\'isica Aplicada i
\`Optica\\ Mart\'i i Franqu\`es 1, 08028 Barcelona, Spain}

\begin{abstract}

This work presents an application that generates real-time holograms to be displayed on a holographic optical tweezers setup; a technique that allows the manipulation of particles in the range from micrometres to nanometres. The software is written in Java, and uses random binary masks to generate the holograms. It allows customization of several parameters that are dependent on the experimental setup, such as the specific characteristics of the device displaying the hologram, or the presence of aberrations. We evaluate the software's performance and conclude that real-time interaction is achieved. We give our experimental results from manipulating 5 $\mu$m microspheres using the program. 

\end{abstract}

\begin{keyword}
Optical tweezers \sep Interactive manipulation \sep Digital Holography \sep Spatial Light Modulators
\PACS 87.80.Cc \sep 87.80.–y \sep 42.40.Jv \sep 42.79.Kr
\end{keyword}

\end{frontmatter}

\begin{flushleft}
PROGRAM SUMMARY

Title of program: HoloTrap\\
Computer for which the program is designed and others on which it has been tested: General computer\\
Operating systems or monitors under which the program has been tested: Windows, Linux\\
Programming language used: Java\\
Memory required to execute with typical data: up to 34 Mb including the Java Virtual Machine\\
No. of bits in a word:8 bits\\
No. of processors used: 1\\
Has the code been vectorized or parallelized?: No\\
No. of bytes in distributed program, including test data, etc.: 1118 KB\\
Distribution format: jar file\\
Nature of physical problem: To calculate and display holograms for generating multiple and dynamic optical tweezers to be reconfigured interactively\\
Method of solution: Fast random binary mask for the simultaneous codification of multiple phase functions into a phase modulation device\\
Typical running time: Up to 10 frames per second\\
Unusual features of the program: None\\
References: The method for calculating holograms can be found in [M. Montes-Usategui, \textit{et al.} Opt. Express, 14 (2006) 2101-2107.]\\
\end{flushleft}

\section{Introduction}
\label{intro}

In this paper we describe an application that interactively generates multiple dynamic holographic optical tweezers. The program allows the user to compute holograms displayed in an optical tweezers setup, resulting in trap patterns that are reconfigurable in real time. Experimental setup factors are application parameters resulting in a completely customizable program.

Optical tweezers are generated by strongly focusing a laser beam, thus creating an optical gradient that traps dielectric particles from micrometric samples down to the nanometric scale \cite{Ashkin}, due to the transfer of light momentum to the trapped particle. This technique has many applications in the manipulation of biological samples \cite{Neumann} because it is harmless and the forces involved in molecular and biological processes (which are in the same range as the forces applied by optical tweezers -about pN) can be measured. 

Holography allows the synthesis of a light wavefront by spatially modifying the amplitude and phase of the beam \cite{raul2}. Knowing how light propagates in the setup allows us to determine the relationship between the field amplitudes in any two planes along the optical train. In this way, we can calculate the hologram that reconstructs a desired light distribution on another plane. The use of digital holography in optical tweezers provides a lot of flexibility in the design of trap patterns. This technique has resulted in the introduction of large arrays of optical traps and three-dimensional control \cite{sinclar}, \cite{grierRoichman}. Moreover, the shape and properties of the beam can be modified to generate non-Gaussian beams such as vortex beams, which are capable of transferring angular momentum to the trapped particle \cite{dholakia}, or non-diffracting beams \cite{tao}.

Spatial Light Modulators (SLMs), which are used to display digital holograms, allow dynamic, computer-controlled modification of the complex transmittance/ reflectance of the device. The relationship between the sample plane and the hologram plane is an inverse Fourier transform, so, in general, the hologram is complex. These devices are constrained to display a set of complex transmittance values, so we should limit our hologram values to those available from the modulator. Algorithms have to be designed to find an optimal solution constraining the hologram to the discrete set of values accessible. These algorithms are based on iterative methods \cite{gerch}, \cite{soifer} or on extensive search procedures \cite{dbs}. Both approaches are time consuming and do not allow real-time interaction with the sample, since they cannot be calculated and displayed as fast as necessary. We recently proposed a method for calculating holograms in order to generate optical tweezers. It is based on the random mask encoding method for multiplexing phase-only filters \cite{Montes}, and is, to our knowledge, the fastest method with 3D control of the trap. This is because it is not iterative and the number of operations involved is lower than in other direct methods, such as the gratings and lenses (or prisms and lenses) method \cite{liesener}. Another advantage of the algorithm is that it does not produce the ghost traps or replicas reported in other methods \cite{grier}.

Other possibilities --such as the generalized phase contrast approach \cite{Rodrigo} or time sharing of the laser beam-- allow real-time interaction with the sample, but are limited to two-dimensional trap patterns and do not allow generation of non-Gaussian beams. The gratings and lenses method has also been used to calculate the desired trap pattern for interactive hologram generation applications\cite{leach}, \cite{whyte}. It has also been proposed direct programming of the graphics card, allowing a faster update rate \cite{reicherter}. Other applications are designed to be used with their commercial setup and do not allow customization \cite{arryx}.

We present software developed to interact with trapped particles in real time. The application calculates and displays the holograms that generate the trap pattern according to the user's commands. In section \ref{hot} we outline the experimental setup, emphasizing the aspects that have to be taken into account in the software design. The implemented algorithm is explained in section \ref{alg}. The developed application, written in Java$^{TM}$ 2 Platform Standard Edition 5.0 is detailed in section \ref{descr}. The sample plane is be visualized on another monitor, using the camera. The camera image can be integrated in the program. We show how to do that in section \ref{cam}. However, the camera driver is proprietary and cannot be attached due to licensing restrictions. The performance of the software and experimental results are given in section \ref{use}. 

\section{Holographic optical tweezers}\label{hot}

In optical trapping, a highly focused laser beam exerts gradient forces on the sample. Typically, an inverted microscope is modified to focus the beam, while still being able to image the sample. Figure 1 shows our experimental setup. The laser is a frequency-doubled Nd:YVO$_4$ laser from Viasho Technologies. The laser beam is expanded and collimated before being reflected by the Spatial Light Modulator, a HoloEye LCR-2500. On reflection, the SLM modulates the phase of the wavefront. The beam size is reduced using an auxiliary telescopic system (lenses L1 and L2 in figure 1, to adapt it to cover the whole of the objective's aperture; which is important for stable trapping \cite{Ashkin}. The beam is fed into the inverted microscope (a Nikon TE2000) through a rear aperture, usually used in fluorescence imaging, and focused in the sample plane by the microscope objective (a Plan Fluor 100X Nikon oil-immersion objective with numerical aperture 1.3).

\begin{figure}[h!]
\centering
\includegraphics[height=5 cm]{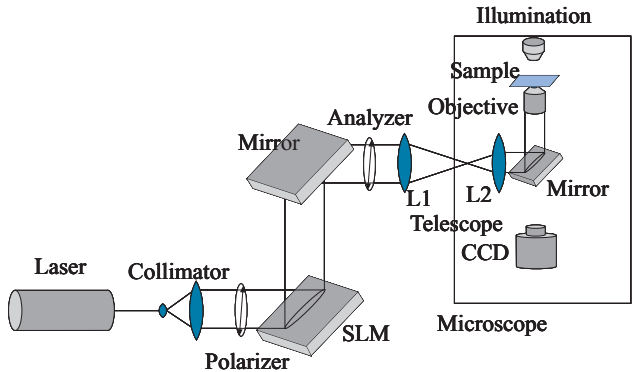}
\caption{Holographic optical tweezers setup}
\end{figure} 

The sample is placed at the focal plane of the objective, so the relationship between the device and the sample is an inverse Fourier transform except for multiplicative phase terms that do not affect our discussion \cite{goodman}:

\begin{equation}
H(u,v)= \iint_{-\infty}^{\infty} g(x,y) e^{-i\frac{2\pi}{\lambda f'}(xu+yv)}dxdy,
\end{equation}

where $H(u,v)$ is the hologram, $g(x,y)$ the trap pattern, $\lambda$ the wavelength of the light and $f'$ the focal length of the objective. The wavelength, the telescope, the modulator and the objective determine the scale factor between our sample plane and the hologram. This is left as a parameter in the application, as explained in section \ref{descr}. 

The introduction of the Spatial Light Modulator allows us to design the shape of the beam by spatially modifying the amplitude and phase of the light distribution in the plane where the modulator is placed. There are different kinds of SLM, such as liquid crystal displays (LCDs) in which the grey level sent to each pixel of the modulator is translated into a change in amplitude and phase of the incident beam at that point, thus achieving spatial control of the light distribution. The modulation also depends on the polarization of the input and output light. It is necessary to know the modulation response for each grey level. This can be achieved by characterizing the device modulation with the polarization conditions \cite{estela} in which it will be used. The most common configuration is phase-only modulation, which has the least amplitude variation.

LCDs are unable to modulate the whole complex plane \cite{raul}. Figure 2 shows the experimental characterization of the LCD we use, a HoloEye LCR-2500. It shows the complex transmittance value that corresponds to each grey level. It is almost a phase modulation from 0 to 2$\pi$, although there is a small amplitude modulation. The hologram values have to be built using the available modulation values. To do this, the minimum Euclidean distance between the phase in each pixel and the available phase values is calculated, and the nearest phase modulation value is used to display the theoretical hologram value.

\begin{figure}[h!]
\centering
\includegraphics[height=8 cm]{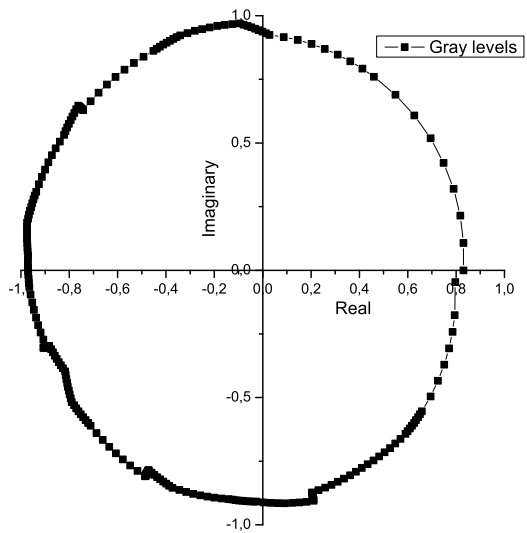}
\caption{Experimental complex modulation values of the SLM as a function of the gray level}
\end{figure} 

To summarize, our synthesized hologram is a grey-level image that results in a local modification of the phase of the incident wavefront, and will generate a given trapping pattern in the focal plane of the objective (where the sample is located).

\section{Fast method of hologram calculation}\label{alg}

In order to generate holograms in real time we have to use direct methods of calculation as opposed to iterative methods. Direct methods consist of generating the hologram from analytic solutions, that is, determining analytically the inverse Fourier transform of the trap pattern taking into account that one is limited in general to phase functions due to modulation constraints. Holographically we have the ability to set the three-dimensional position of each trap. A trap centred at $(a,b)$ can be described as $g(x,y)= \delta(x-a,y-b)$. The hologram that generates this distribution is its Fourier transform, that is:

\begin{eqnarray}\label{eqlinph}
H_D(u,v) = \mathcal{F}^{-1}(g(x,y)) = \exp\big(i\frac{2\pi}{\lambda f}(u\cdot a+v\cdot b)\big).
\end{eqnarray}

If the trap is focused at a depth $z$ from the focal plane, the required function is a quadratic phase term such as:

\begin{equation}\label{eqquad}
H_Z(u,v) = \exp\big(i\frac{2\pi}{\lambda z}(u^2+v^2)\big),
\end{equation}

whose Fourier transform is another quadratic phase function.

To generate a vortex, which can transfer angular momentum to the trapped particle \cite{curtis}, the following phase function is needed:

\begin{equation}\label{vortex}
H_V(u,v) = \exp\big(i \cdot l \tan^-1(\frac{v}{u})\big),
\end{equation}

this function modifies the wavefront to convert it to a Laguerre-Gaussian mode, which carries angular momentum. The quantity $l$ is called the topological charge and is related to the orbital angular momentum of each photon by $L = l\hbar$.

In these examples the solution is a pure phase function, so codifying it using phase-only modulation is straightforward: simply choose the closest phase given by the device. However, if $N$ traps are required, the hologram is a sum of as many phases as traps displayed, resulting in a complex function that cannot be directly displayed with a phase-only modulation:

\begin{equation}
\label{sum_ph}
H(u,v)=\sum_{k=0}^N\big(H_{Dk}+H_{Lk}+H_{Vk}\big) \neq \exp\big(i\phi(u,v)\big).
\end{equation}

The method for the codification of the hologram cannot be time-consuming if we require it to be implemented in real time. Our approach, more detailed in \cite{Montes}, defines as many different domains $I_k$ of the modulator as traps to be displayed. Each domain consists of a set of modulator pixels that shows a phase function. In this way, each set is in charge of generating a single trap. The hologram (equation \ref{eqholo}) consists in the multiplication of the phase functions, $H_k(u,v)$, (as in equations \ref{eqlinph} and \ref{eqquad}) by spatially disjoint binary masks, $m_k(u,v)$ - the set of pixels of the domains $I_k$. 

\begin{eqnarray}\label{eqholo}
H(u,v) = \sum_{k=0}^{N}m_k(u,v)\cdot H_k(u,v),
\end{eqnarray}

where

\begin{displaymath}
m_k(u,v) = \left\{ \begin{array}{ll}
1 & \textrm{if $(u,v) \in I_k$}\\
0 & \textrm{elsewhere.}\\
\end{array} \right.
\end{displaymath}

The domains $I_k$ do not overlap, and together they cover the whole modulator. For example, we can generate the domains by randomly deciding which pixels will belong to each trap. This is a good choice since the mask that defines every sub-hologram is not a geometric function: the convolution of the shape of the trap with the Fourier transform of the mask would result in noise distributed through the resulting plane \cite{Montes}. As can be seen, the solution is as fast as generating the $I_k$ domains each time a trap is added or deleted, and computing the arguments of the complex exponentials $H_k(u,v)$ to display the hologram. Figure 3 shows an example of a hologram in which half of the pixels display a linear phase function and the other half a quadratic phase. The resulting light distribution would be two different traps placed off-centre, at two different depths.

\begin{figure}[h!]
\centering
\includegraphics[height=5 cm]{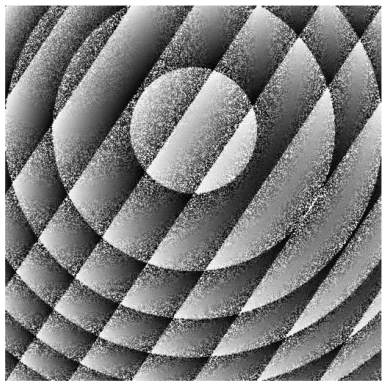}
\caption{Hologram calculated using the random binary masks method}
\end{figure} 

This method is --to our knowledge-- the fastest way to generate simple trap patterns. The most common fast method for the generation of optical tweezers (the gratings and lenses method) consists of generating the analytic hologram from equation \ref{sum_ph} and discarding the amplitude information. This method is slower than our random binary masks method because the calculation time increases with the number of traps and requires the computation of trigonometric functions \cite{liesener}. Due to the discarded amplitude information, the intensity distribution among the traps may be different from that expected. Another advantage of our random binary masks method is that the trap pattern generated does not present replicas \cite{grier}. While other methods tend to concentrate the energy not located in the traps in bright spots, resulting in false traps, the random binary masks method distributes the remaining energy in noise, due to the convolution with each random mask. The downside is that the efficiency of the traps is lower than that achieved with other methods. 

\section{Description of the program}\label{descr}

The software we present calculates and displays holograms to generate optical tweezers allowing real-time interaction with samples. Each change in the number or position of the trap requires recalculation of the hologram. The program responds quickly enough to provide close to video-rate feedback from the sample. 

The reason for using Java is that development costs are low. Moreover Java allows easy generation of the Graphical User Interface (GUI) and easy integration with C++ generated dlls; the most generalized hardware driver distribution method. Another advantage of using Java is simple remote control of the experiment over the Internet. If the computer controlling the camera acts as a web server, you just have to transform the program generated into a servlet and use the Remote Method Invocation (RMI) classes.

The source code is distributed into three different classes. The first class, TRBase, generates the GUI and handles the events related to the input parameters. It also initiates the second class, PanelCoord, the panel in which the user clicks and drags to generate and move a trap, and so this class monitors these mouse events and calls to the third class, FrameHolo. This third class calculates and displays the hologram using the mouse coordinates and the input parameters. The documentation of the application, in which the different classes and implemented methods are detailed, can be found in the folder $/html$ zipped within the application jar file. This documentation can be also found in our website \cite{web}.

\subsection{Graphical User Interface}

A screen capture of the GUI can be found in figure 4. This program allows user control of several variables and initial data: 

\begin{itemize}
\item The hologram size, in pixels. If the size is set to 1024$\times$768, which is our SLM full resolution, the hologram is calculated with half the number of and zoomed to fill the modulator, reducing computation time.

\item The scale factor between the Spatial Light Modulator plane and the visualization plane. The scale can be modified by changing the Row and Column factors. This scale factor can be found experimentally, or deduced from the different experimental parameters: telescope, SLM, objective, and field of view \cite{jopa}.

\item The Init button asks for a file containing a precalculated map of the phase modulation and a phase aberration correction (see section \ref{calc}). In our case, the aberration is a distortion of the wavefront due to the curvature of the modulator surface. There is an example of a phase-only function map and a null aberration correction attached in the \textit{.jar} file to check the required format. To run the application using these two ideal condition files, after pressing the Init button, just press OK on the dialog box "Use the default aberration and modulation files". Each time a hologram is generated, the correction is added and then the nearest grey value is assigned using the precalculated map. 

\item A selector allows you to choose the manipulated trap if more than one trap is generated. The selected trap is indicated by a red circle, whereas the unselected traps are in green.

\item A slider allows you to modify the trap depth, from -5$\mu$m to 5$\mu$m. The "Depth Factor" text field allows you to modify the available depth range.

\item By changing the integer in the "lvortex" text field (see figure 4) an optical vortex carrying angular momentum is generated by adding a vortex phase function (equation \ref{vortex}) with the specified topological charge.

\item The Delete trap button deletes the selected trap. This involves a reconfiguration of the random binary masks, which have to be recalculated.

\item The hologram is calculated by the method selected in the Radio Button. The default calculation method is random binary masks, but gratings and lenses is also available.
\end{itemize}

\begin{figure}[h!]
\centering
\includegraphics[height=8 cm]{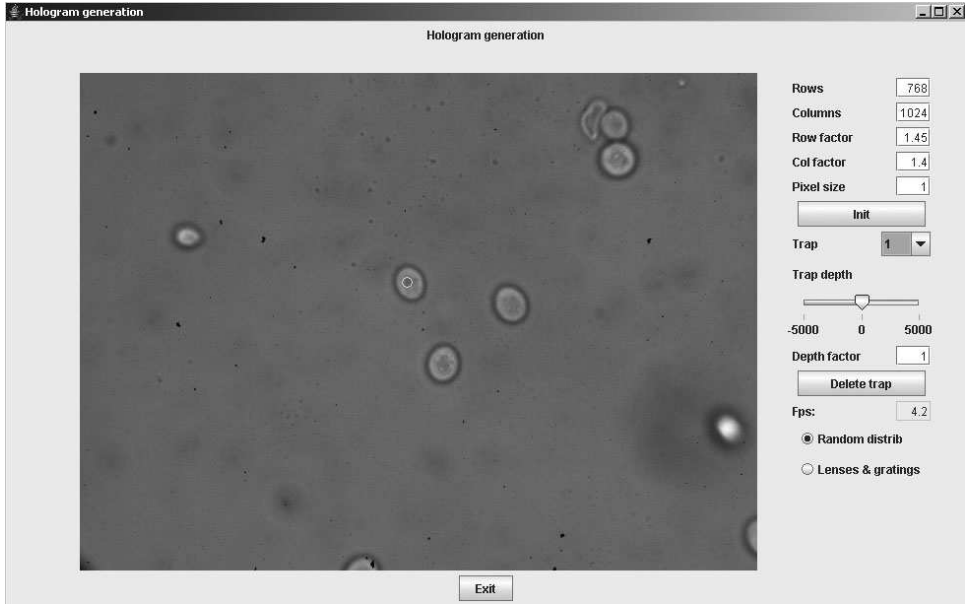}
\caption{Application user interface}
\end{figure} 

\subsection{Calculating and displaying a hologram}\label{calc}

This section details the computational process of generating a hologram, see figure 5. First, the application is initiated and the user enters the desired parameters (such as hologram size or scale factors). The central panel in the application controls the event handling of the user interaction. When the mouse is clicked on the panel, the mouse coordinates are obtained. The random mask is calculated, resulting in the whole modulator, because there is one single trap in this first step. With the mouse coordinates, a phase grating corresponding to the trap is calculated (equation \ref{eqlinph}) and the hologram is displayed. Each time another trap is added, the random binary masks have to be calculated and then each set of pixels show the corresponding phase function. If the mouse is dragged or the depth slider is moved, the coordinates of the selected trap change. A change in the coordinates of a single trap means that only the pixels of the mask corresponding to that trap have to be recalculated.

\begin{figure}[h!]
\centering
\includegraphics[height=10 cm]{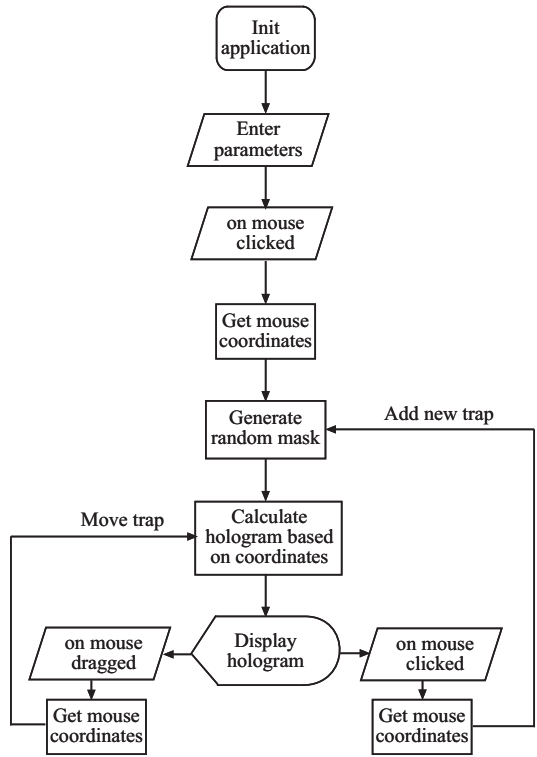}
\caption{Program Flowchart}
\end{figure} 

To generate traps in real time, the algorithm has to be fast, but there are also other considerations that affect the performance of the process. We have optimized the process of adaptation of the hologram to the modulation values by generating a map of the correspondence between all the possible phase values between 0 and 2$\pi$ and the nearest phase given by a grey level. In general this is not a linear relationship. The phase value assigned to each grey level is stored in a file that the program reads as an initial parameter. An example of an ideal phase assignment can be found in the $map\_ideal.txt$ file distributed in the jar. Once the analytical phase value is calculated, the map provides the grey level to be displayed. An incorrect assignment can cause variations in the reconstructed trap pattern. 

Another experimental issue that affects the calculation time is the possible existence of aberrations in the optical system, which can be corrected with the SLM when generating the trap pattern. In our case, the modulator reflected wavefront is distorted because the device is not flat. We can correct this aberration by adding a fixed phase pattern to each hologram. As an example, the file $phaberr\_1024$x$768.txt$ is a null aberration correction attached in the jar file, that shows the format of the aberration file for the specified hologram size. 

In order to ensure a fast response of the displayed hologram, two main factors have to be taken into account. First, the hologram generation has to be as fast as possible, including algorithm calculation, addition of the aberration correction and adaptation to the modulation. Second, speed of access to the graphic hardware has to be ensured. Our approach is to take advantage of the $VolatileImage$ class in the Java SDK. The hologram is stored as a hardware-accelerated off-screen image, in such a way that rendering operations are accelerated through the graphics card. Thus, hologram is displayed without using the CPU. This class parallelizes the display of the hologram and the calculation processes, with the CPU performing the calculation. 

\subsection{Camera control}\label{cam}

The image of the sample can be displayed on another monitor to control manipulation. Our program is enhanced if the camera image is incorporated into the interactive interface, although it can be used with the image separated from it. In this section we explain how we integrated our video stream management, as a guide for users on how to embed their own. We used a QICam Fast 1934 from QImaging Corp. \cite{qim} camera. It is not compliant with the IIDC Digital Camera Specification (DCAM), which is the standard protocol FireWire cameras should follow, so the SDK provided by the manufacturer had to be used. This is a drawback to distributing the camera-integrated version of the program, and so a version without a camera accompanies the paper. If a DCAM-compliant camera is used, the Java API for FireWire $jlibdc1394$ \cite{jlibdc1394} can be incorporated into the program instead of the camera SDK, making it suitable for all DCAM-compliant cameras. 

The Qimaging libraries have to be used with a C++ compiler, so the Java Native Interface (JNI) class \cite{jni} is needed to embed the camera library into the Java structure. JNI is a common trick for gaining compatibility with native methods across a Java virtual machine. We need the following native functions: 

\begin{flushleft}
\begin{verse}
public native int initCamera(); (Detects the camera)\\ 
public native int initGrab(); (Initiates the recording)\\ 
public native int doGrab(byte[] pix); (Saves the image into a pixel array)\\
public native int StopGrab(); (Stops recording)\\
\end{verse}
\end{flushleft}

Each native Java method has its corresponding function in C++. The process of calling from a Java program code contained in the proprietary library is \cite{jni}:

\begin{itemize}
\item[-] Declaration of the native methods in the Java application, in our case the methods listed above. 
\item[-] The loading of the library containing the native code implementation, by calling the function \textit{System.load("JNIQCam.dll")}, where 'JNI\-QCam.dll' is our generated library name (even it does not exist yet). The Java application has to be compiled at this point without being executed. This library is not the proprietary library, but one generated by the user, defining what each native method does.
\item[-] Generation of the header (.h) file that contains the interface assigning the Java methods to the C native functions. As an example, the functions are defined in this header as:

\begin{verse}
JNIEXPORT jint JNICALL Java$\_$initCamera(JNIEnv *, jobject);\\
JNIEXPORT jint JNICALL Java$\_$initGrab(JNIEnv *, \mbox{jobject});\\
JNIEXPORT jint JNICALL Java$\_$doGrab(JNIEnv *, \mbox{jobject}, jbyteArray);\\
JNIEXPORT jint JNICALL Java$\_$StopGrab(JNIEnv *, \mbox{jobject});\\
\end{verse}

This file is the communication channel between the two languages.

\item[-] Creation of the C++ functions. The library (JNIQCam.dll) has to contain the C++ source calling to the camera library. As an example, our C++ method that disconnects the camera is:

\begin{verse}
JNIEXPORT jint JNICALL Java$\_$tr$\_$StopGrab(JNIEnv *, \mbox{jobject)$\{$}\\
	delete [] pixels;\\
	if(hCamera != NULL)\\
	$\{$ \\
		QCam$\_$CloseCamera( hCamera );\\
	$\}$\\

	QCam$\_$ReleaseDriver();\\
    return 0;\\
$\}$
\end{verse}

In this example we free the image memory through the \textit{delete} order. The calling to \textit{CloseCamera(hCamera)} frees the \textit{hCamera} object, \textit{hCamera} is the object initiated in the method \textit{initCamera}, which contains the camera properties and prevents other applications accessing the camera. Next, the camera driver is released with the command \textit{ReleaseDriver()}.

\item[-] Compilation and execution of the code. 

\end{itemize}

If the user had the same camera, a .dll file should be generated and the commented lines in the .java source, marked as "$//$Comment if there is no QICam available", should be uncommented.

\section{Performance of the software}\label{use}

Figure 4 shows a screen capture of the program. The tests were carried out on a Pentium IV HT, 3.2 Ghz, with lite versions of the program, where not all the options were available. These lite versions can be obtained from our website \cite{web}.

The speed of the software when generating holograms in response to a mouse drag is about 10 fps (frames per second). This value measures the number of holograms displayed per second on the modulator. The full resolution sized holograms are achieved by calculating holograms of 512$\times$384 pixels and resizing them into 1024$\times$768 pixels. The adaptation mapping that we have created from the experimental phase modulation values does not affect the speed of the hologram generation. In contrast, the inclusion of the aberration correction affects slightly the performance by decreasing the hologram calculation speed. The time response does not increase with the number of traps, because the number of pixels the phases have to be computed for (the number of pixels defining each mask) decreases as the number of traps increases. 

Figure 6 shows screen shots of experimental manipulation of polystyrene beads of 5$\mu$m diameter. A first microsphere is captured and dragged close to another, which is trapped and moved with a second trap. 

\begin{figure}[h!]
\centering
\subfigure[]{
\includegraphics[height=5 cm]{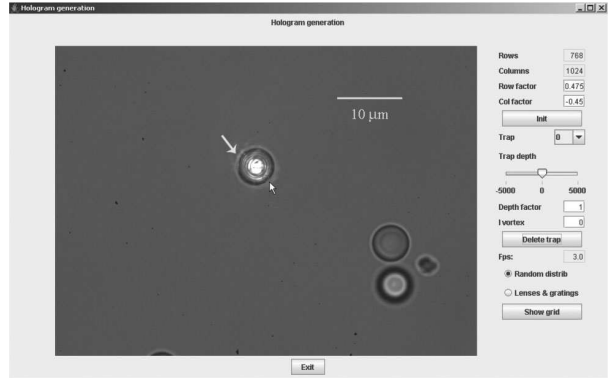}}
\subfigure[]{
\includegraphics[height= 5 cm]{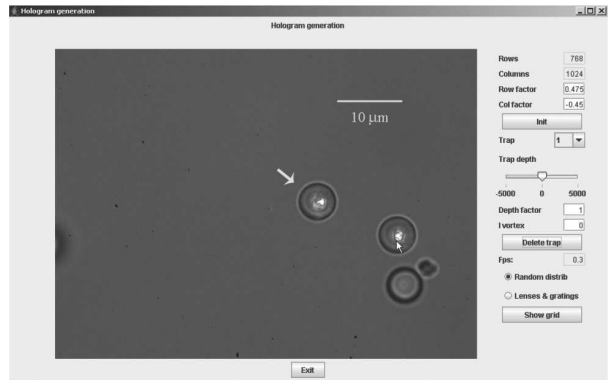}}
\subfigure[]{
\includegraphics[height= 5 cm]{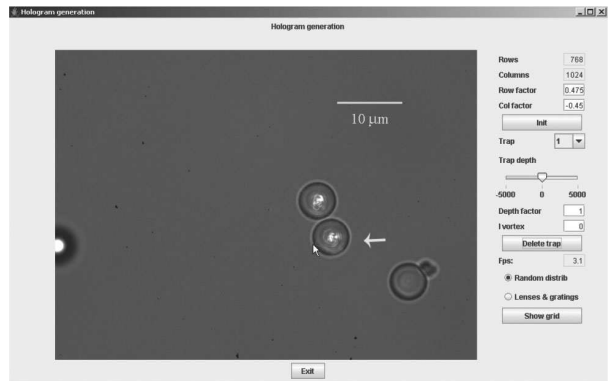}}
\caption{Screen shots showing the experimental manipulation of polystyrene beads, 5$\mu$m diameter}
\end{figure}

\section{Concluding remarks}

We have presented an application for calculating and displaying holograms in real time to generate multiple reconfigurable optical tweezers. The application allows the user to generate, delete or modify optical traps interactively. We used the random binary masks method because of its speed. The software takes into account different parameters given by the experimental setup, and so it is not limited to a single configuration. The different options have been detailed, including the adaptation to the modulation and the correction of possible aberrations. The scale factor and the hologram size can also be modified. Some strategies for accelerating hologram calculation and display are explained. A second version of the program takes advantage of the proprietary libraries of the camera used in order to embed the image provided by the camera and the program. The viability of the software is comparable to that of other applications in the literature. We include an example of optical manipulation using this program. In future work we would like to make the software compatible with IICAM-compliant FireWire cameras. 

ACKNOWLEDGMENTS

This work has been funded by the Spanish Ministry of Education and Science, under grants FIS2004-03450 and NAN2004-09348-C04-03.

\end{document}